\documentclass[10pt,a4paper,twocolumn]{article}
\usepackage{times}
\usepackage{psfig}
\def\thewidth{\columnwidth}
\def\halffig{3.5cm}
\let\OLDLARGE=\LARGE
\renewcommand\LARGE{\OLDLARGE\bfseries}
\usepackage{caption}

\usepackage{cite}%other possibilities are cite,overcite,drftcite

\renewenvironment{itemize}
 {\list{$\bullet$}{\setlength{\parsep}{0mm}\setlength{\itemindent}{0mm}%
\setlength{\topsep}{1mm}\setlength{\leftmargin}{7mm}}}
{\endlist}
\def\mybiblio{\small %!!!!
\itemsep=0pt         %!!!!
\parskip=0pt}        %!!!!
\def\startfigure{\begin{figure}} %!!!!
\begin{document}

\title{Curvature of Co-Links
Uncovers Hidden Thematic Layers in the World Wide Web}
\author{Jean-Pierre Eckmann\footnote{D\'ept.~de Physique Th\'eorique,
Universit\'e de Gen\`eve,
CH-1211 Gen\`eve 4, Switzerland and
Section de Math\'ematiques, Universit\'e de Gen\`eve, CH-1211
 Gen\`eve 4, Switzerland. }, \ Elisha
Moses\footnote{Department of Physics of Complex Systems, The Weizmann
Institute of Science, Rehovot 76100, Israel}}
\maketitle

\begin{abstract}
{\bf Beyond the in\-form\-ation stored in pages of the World Wide
Web, novel types of ``meta-in\-form\-ation'' 
are created when they connect to each
other. This in\-form\-ation is a collective effect of independent users
writing and linking pages, hidden from the casual user. Accessing it
and understanding the inter-relation of connectivity and content in the WWW is
a challenging problem \cite{1,12,10,review}.
We demonstrate here how thematic relationships can be located precisely by
looking only at the graph of hyperlinks, gleaning content and context from the
Web without having to read what is in the pages. We begin by noting that reciprocal links
(co-links) between pages signal a mutual recognition of authors, and
then focus on
triangles containing such links, since triangles indicate a transitive
relation. The importance of triangles is quantified by the clustering
coefficient \cite{Watts}~which we interpret as a curvature
\cite{Gromov,Bridson-Haefliger}. This defines a Web-landscape whose connected
regions of high curvature characterize a common topic. We show experimentally
that reciprocity and curvature, when combined,
accurately capture this meta-in\-form\-ation for
a wide variety of topics. As an example of future directions we analyze the
neural network of {\textit {C. elegans}} \cite{White, Wood}, using the same methods.
}\end{abstract}

The Web is a graph \cite{7,8} that is continuously being expanded by an
enormous number of {\em independent} agents. Millions of users add pages in an
uncoordinated way, but in doing so they must obey clear rules on how to address other
pages \cite{3}. 
Beyond the information placed by the users directly in individual pages, their
unified action creates a {\em cooperative meta-in\-formation} whose locus is in
the connectivity of the network.  By meta-information we mean here information that is
based solely on the graph structure of the WWW, and which reveals
strong contextual grouping. 
This type of information should be
contrasted to that of a very informative but secluded page, that is an archive
of in\-form\-ation (a ``wise hermit''). Further on, we will compare our new
meta-information to earlier work.

The billions of pages and links of the WWW create a maze of confusing and
almost unmanageable complexity. To navigate through
it and to find pages of interest,
users rely on the help of search engines \cite{9}. These send out
small, automated programs (robots) that roam the Web, retrieve links
and 
note keywords as they create a mirror image of the Web that can be stored
within one room of storage data, ready for easy access. The Web remains
intelligible to its human users precisely because it is constantly analyzed and
monitored by these automatic agents. Robot activity dominates Web traffic and
accounts for up to 70\% of the total queries made at certain sites we studied.
Because of the breadth of their search, robots obtain a clearer view of the
Web than the common user.

It is therefore useful to adopt the perspective of robots,
without having to define meaning or content, looking only
at the identification tag of pages (URL addresses) and their links \cite{11}.
To avoid the difficulty of having to know the complete Web graph we
use robots which explore
some locally  visible pieces that can be found by starting from a given page
and progressing  along its links. An analogy may be drawn with classical
neuro\-biology \cite{Cajal}, where a neuron takes up dye which then traces out
its connections.

Having defined our aim of revealing meta-in\-form\-ation in the form of hidden
thematic structures, we implement it by a novel combination of a number of key
ideas: {\bf Clustering, co-links, triangles, and curvature}.
The clustering regroups a ``home'' page with the pages in its sub-directories.
The co-links are then the result of independent agents linking their pages.
Triangles capture transitivity, which we measure by the associated notion of
curvature. Using it, we show that practically all the connective
in\-form\-ation depends on co-links. Thus, our method reveals
a new geometric view of the locus of
in\-form\-ation in networks, which is not
``put there'' but ``happens'' as a collective effect.

In detail, we proceed as follows:
The first step of our method is  {\em clustering} \cite{Cluster}. Once a
certain number of pages have been found by our robot, we regroup them into
a node by
lumping together a ``home'' page with the pages in its
sub-directories. This is useful because:
\begin{itemize}
\item{In\-form\-ation within a given home site belongs to one contextual
heading.}
\item{The size of the graph is reduced by a factor of $10^2$--$10^3$.}
\item{Connective links that arise from physical proximity are eliminated,
leaving the more relevant remote links.}
\end{itemize}

Working henceforth exclusively with the clustered graph (like the one on the
right of Fig.~1), we introduce the pivotal notion of {\bf co-link} for a
reciprocal connection between two nodes $A$ and $B$ ($A$ points to $B$ {\em
and} $B$ points to $A$, the red link in Fig.~1). They are then deemed to be
``congruent'' (friends, or members in the same interest group).

\startfigure
%\centerline{\psfig{figure=fig1new2.ps,height=6 cm}}
\centerline{\psfig{figure=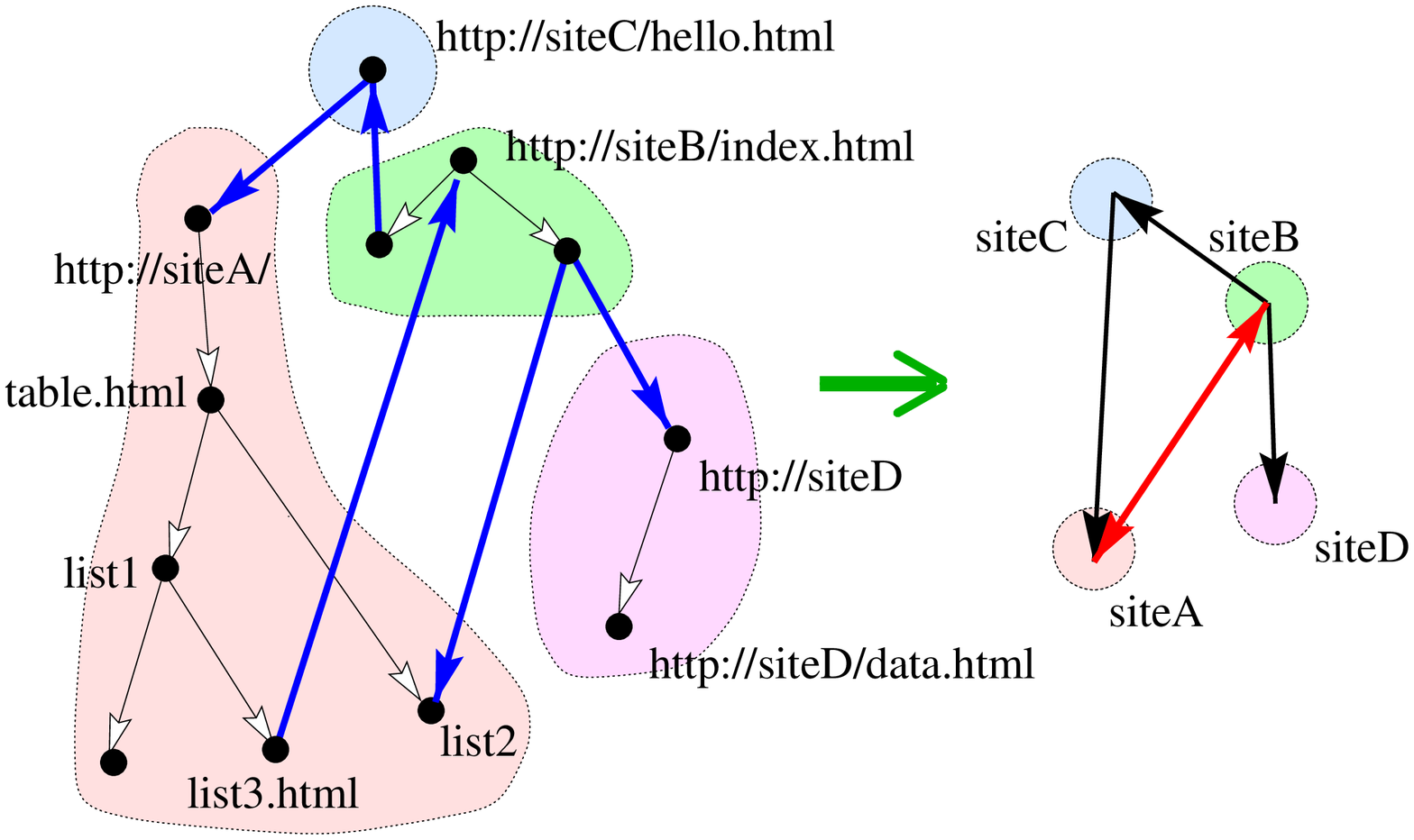,width=\columnwidth}}
\caption{The reduction of a set of
pages and links by clustering. On the left, local links are black and
remote links are blue. On the right, black links are One-way and the red
one is a co-link. Clustering clarifies connectivity because it allows the
links to go through {\em different} pages within the nodes.}
\end{figure}

Co-links are special because, while I can point to your page I cannot (in
general) cause you to point to mine. Congruence therefore indicates an
awareness by both nodes of the other's existence and content, and a recognition
of the other's value to their own interests. Since congruence involves mutual
recognition, we expect it to be a transitive property i.e., if sites $A$ and
$B$ are congruent, and $B$ and $C$ are congruent, then $A$ and $C$ will be
congruent with high probability, forming a {\em triangle}. Such triangles
signal strong {\em co-operative content}, and would be extremely 
rare in any large random graph
\cite{Bollobas}. To quantify the aggregation of triangles with congruent edges,
we define the {\em local curvature} at a node $n$ by $c_n=2t_n/((v_n-
1)v_n)$. This quantity was termed the clustering coefficient in
\cite{Watts}. Here, $t_n$ is the number of triangles containing $n$ as a corner,
$v_n$ is the number of links leaving $n$ (the valence) and
$(v_n-1)v_n/2$ is
the maximal number of possible triangles. We ignore in this the directionality
of One-way links. If the graph is a tree, there are no triangles, and $c_n=0$
everywhere, while for any complete graph (where all nodes are mutually
connected) $c_n = 1$ everywhere.

\startfigure %Fig2
%\centerline{\hbox{\psfig{figure=sphere.ps,width=\halffig,angle=270}}
\centerline{\hbox{\psfig{figure=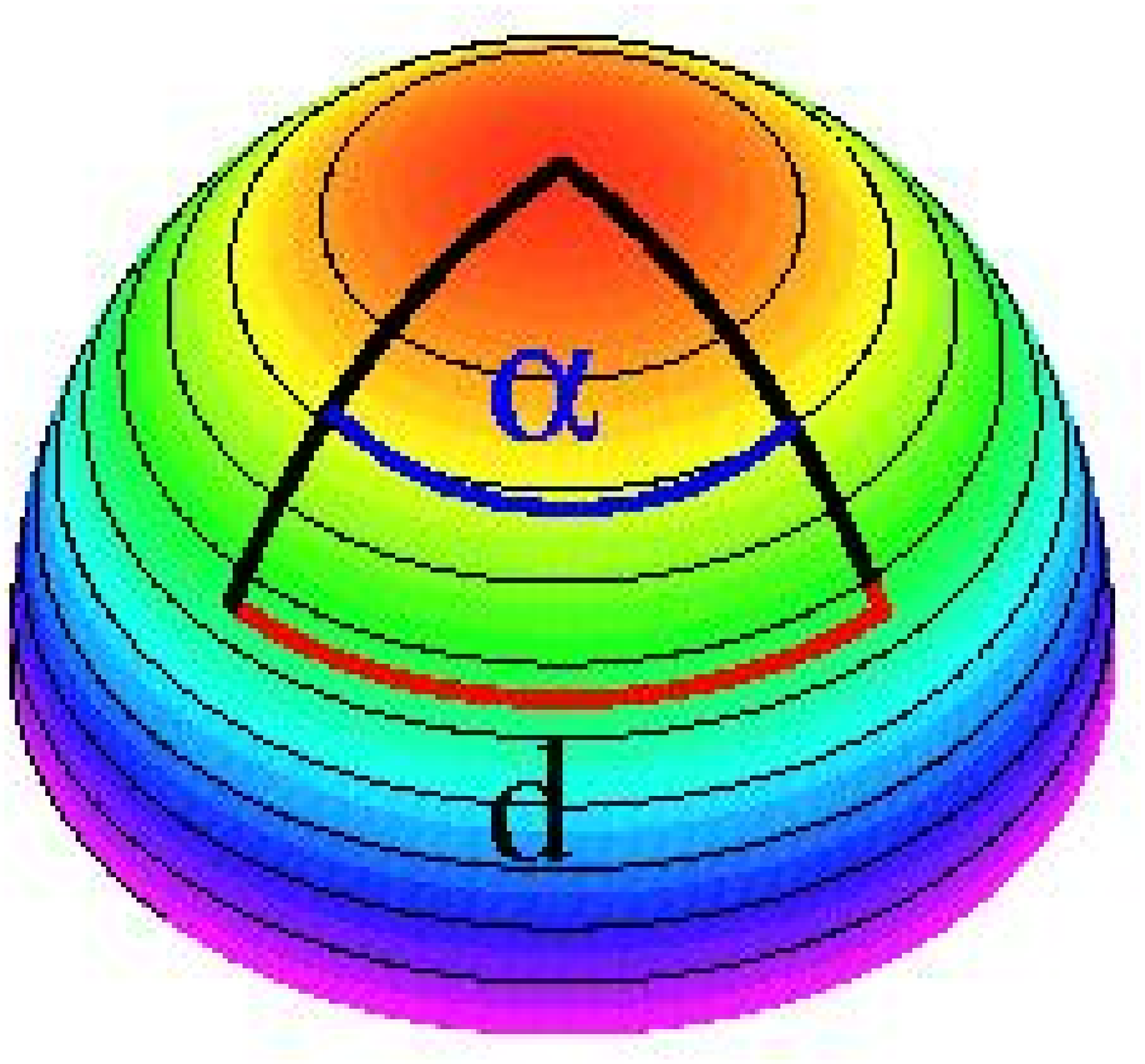,width=\halffig,angle=270}}
%\hbox{\psfig{figure=hyp3.ps,width=\halffig,angle=270}}} 
\hbox{\psfig{figure=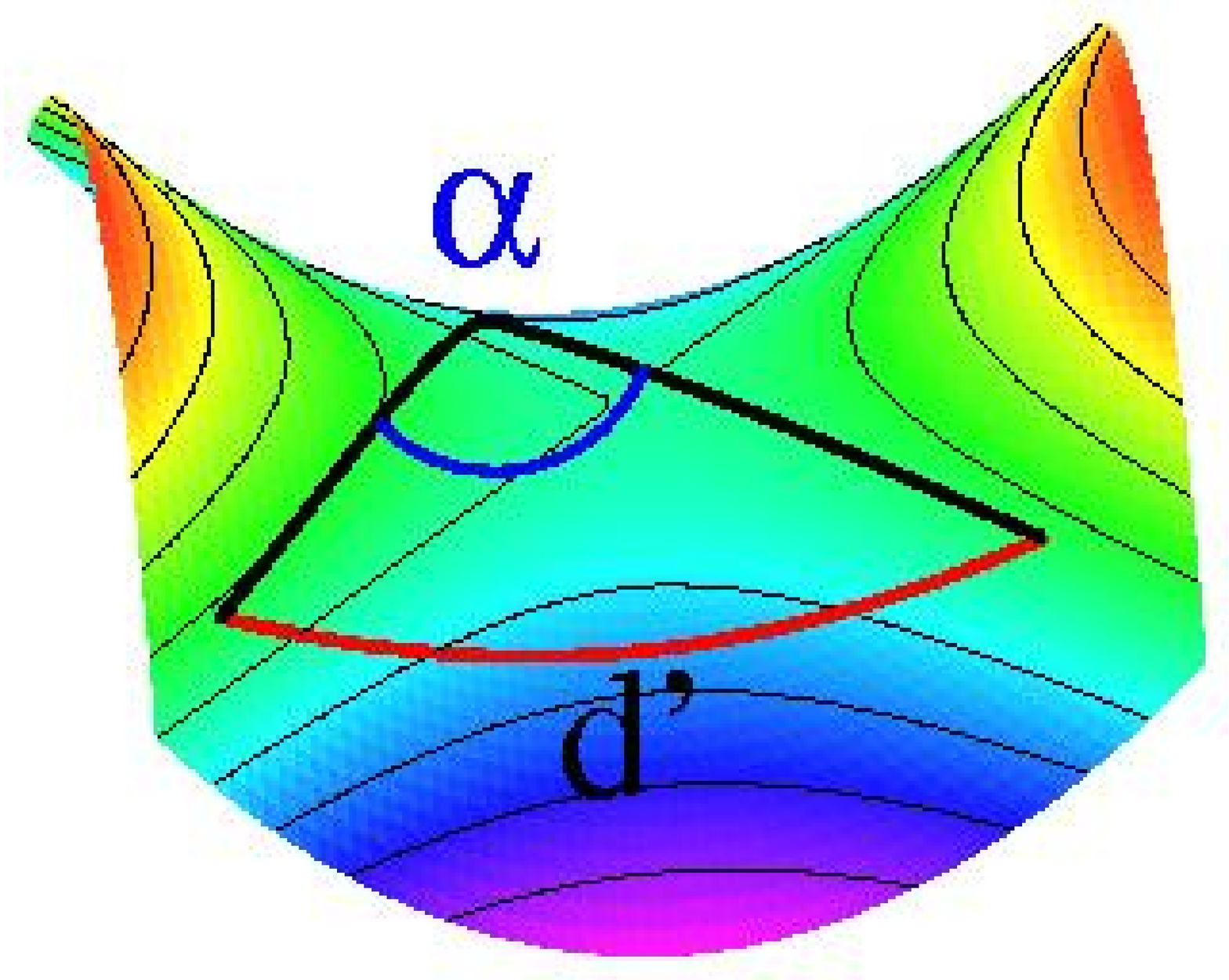,width=\halffig,angle=270}}} 
\caption{Illustration of the law of cosines on a sphere (positive
curvature) and a hyperboloid (negative curvature). The triangles have
sides 1 and 1 and the same angle $\alpha $, but the distances between
the ends are different: $d<d'$.}
\end{figure}

To see that $c_n$ is indeed a quantity like curvature, we regard the simple
geometric picture of Fig.~2: One way to detect that we are on a
curved surface using local measurements is to walk a unit distance on a
straight line in two different directions separated by a constant angle
$\alpha$. Connecting the two endpoints completes a triangle, and the length of
the third edge is determined by the curvature. On surfaces
this relation of triangles to curvature is a natural generalization of the law
of cosines, relating the angle at the apex and the three sides. For 
general metric spaces, such as infinite graphs,
this relation is the basis of a mathematical
definition of curvature \cite{Bridson-Haefliger}
obtained by comparing triangles in a continuous metric to their counterparts
embedded in standard manifolds.

The average number of triangles $c_n$ at a node $n$ is related to
the average distance between any two of its nearest neighbors (say
$n'$ and $n''$). This distance is
measured by counting links in the shortest path from one
node to another, namely $n'$ and $n''$.
Because for co-links this is either $1$ ($n'$ and $n''$ directly
connected, triangle exists) or $2$ (no triangle, need to go through node $n$),
$c_n= 2-\langle r(n',n'') \rangle$, where $r$ is the distance. The
average $\langle\cdot\rangle $ is
taken over all pairs of nearest neighbors, so that the average triangle has sides
1, 1, $\langle r \rangle$. In principle, for One-way links the distance $r$ may
be a large number and can range all the way to infinity (no connection), but we
shall still use $c_n$ as an upper bound on the connectivity.

\def\thescale{ hscale=60 voffset=250 hoffset=-50 vscale=60 angle=270}
\startfigure %FIg3
%\centerline{\psfig{figure=fig2a.ps,width=\thewidth,angle=270}}
%\centerline{\psfig{figure=fig2b.ps,width=\thewidth,angle=270}}
\centerline{\psfig{figure=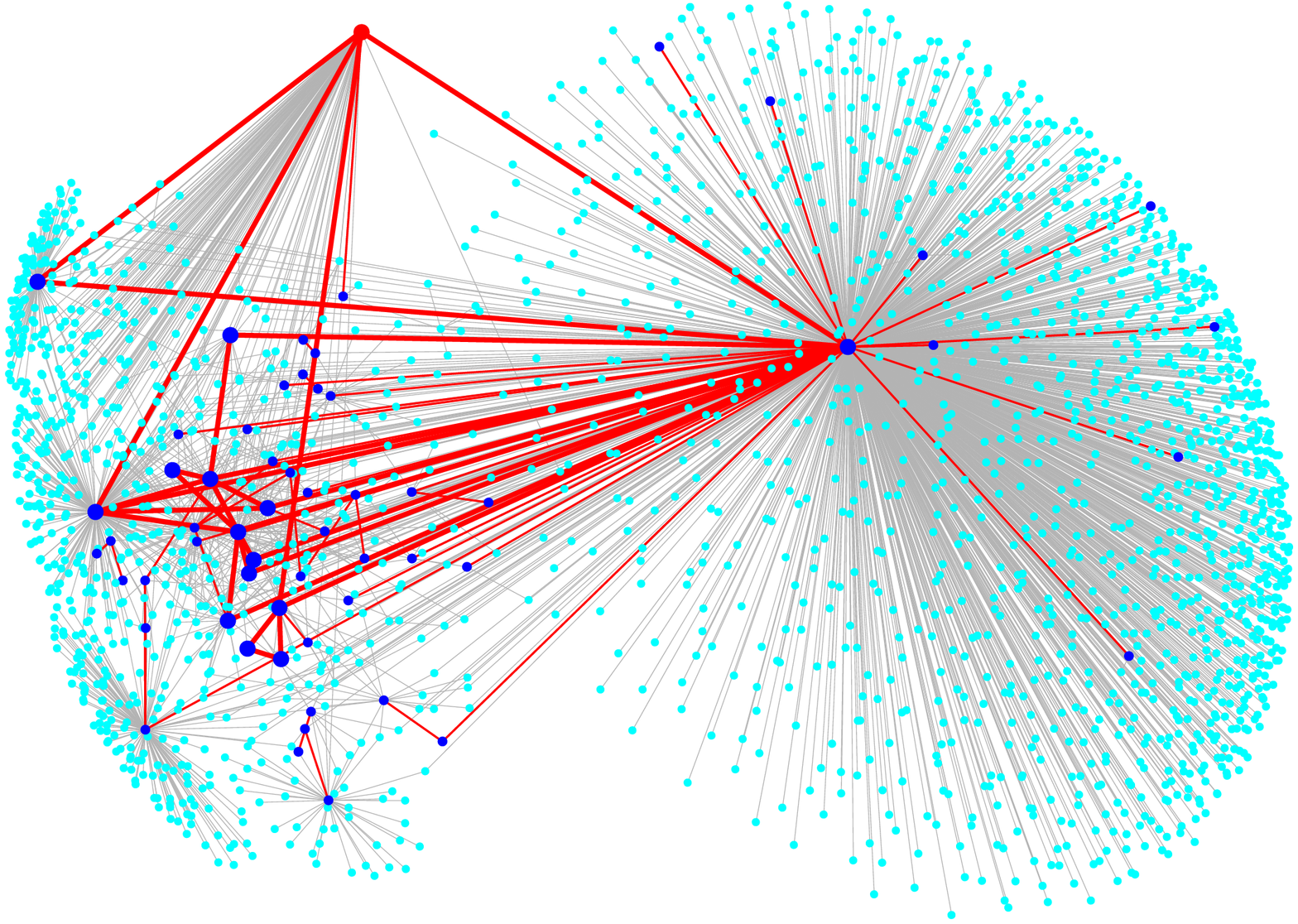,width=\thewidth,angle=270}}
\centerline{\psfig{figure=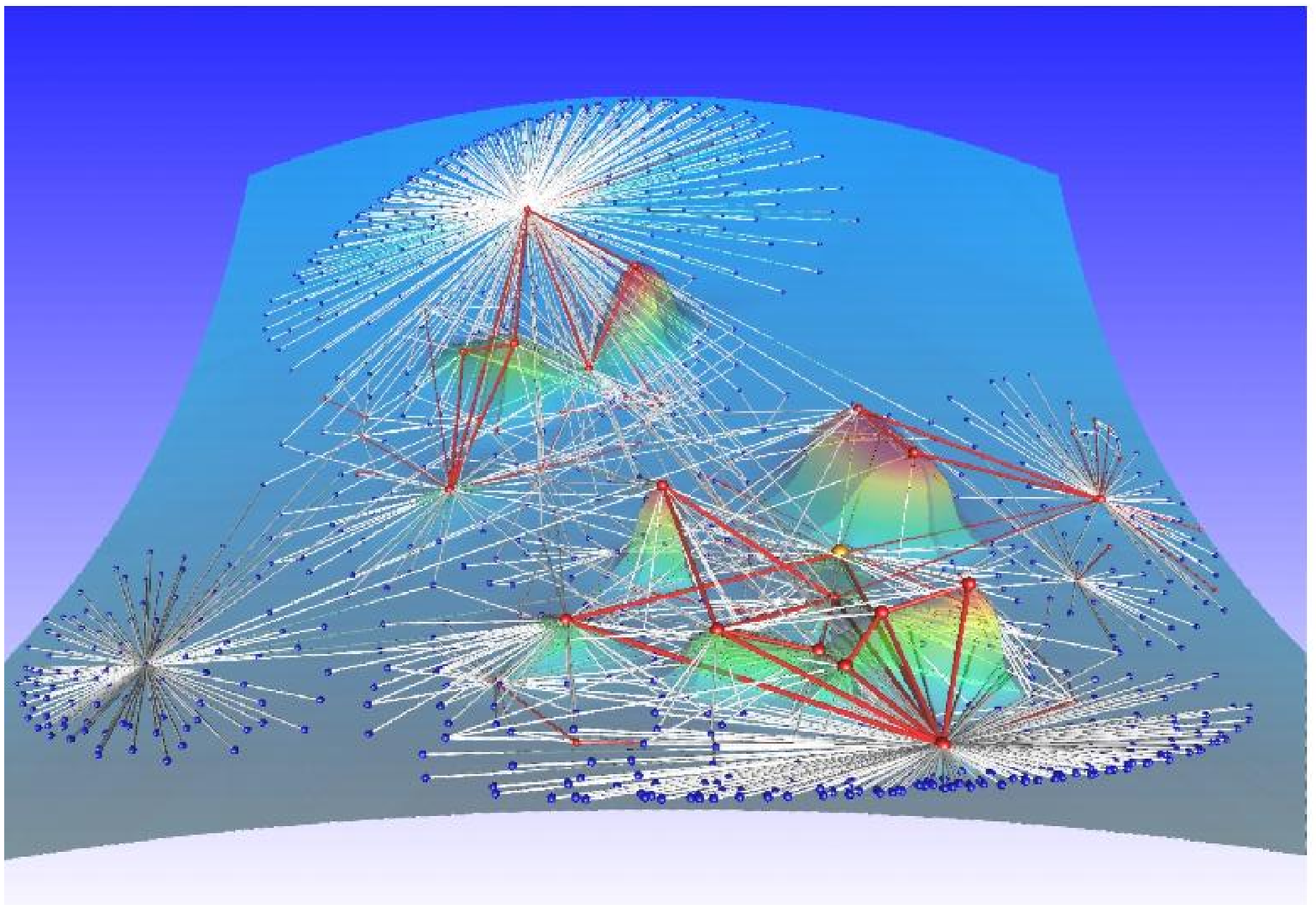,width=\thewidth,angle=90}}
\caption{
Top: 2,125 nodes and 2,755 links of a site
specializing in
tango
(specifically the music of Piazzolla).
The co-links are shown in
red, congruence triangles are
enhanced and one-way links are black. The anchor is red. The sparsity and high
connectivity
of the ``Congruence'' graph are striking.
Bottom:A view of the
curvature for the $947$ nodes and the 1,307 links of a notorious swiss
revisionist (antisemitic) site. Height above the blue surface is proportional to curvature
(of
congruence triangles). The One-way graph is clearly less connected, its links
are more likely to be dangling ends, and it is generally dominated by the
appearance of hubs.}
\end{figure}
We are not aware of a definition similar to the one of
\cite{Gromov,Bridson-Haefliger} for
finite graphs with discrete metric, but it is still useful to view $c_n$ as
the curvature of triangles with apex
at $n$ (though $c_n$ is normalized to take values between $0$ and $1$), with
$c_n$ being perhaps closer to a dimension measurement \cite{99}. Equipped with
a metric (minimal number of hops between nodes) and a curvature, the graph of the WWW can
be visualized geometrically. We shall say that it looks like a hyperboloid when
it is tree-like, with exponential separation of branches that ``fan
out'' \cite{1}. See Fig.~3.
 In contrast, a highly connected region looks more like a sphere that is
``closed'' on itself. Alternately, a random walk by the robot in the
graph will tend to get trapped somewhat longer in highly curved regions.

Triangles tend to aggregate, creating interest groups of widely
varying size (in terms of the number of members), but of small
diameter. In general, the border of interest groups is expected to be
a sharp interface, but sometimes different 
interest groups will connect one to another, typically via a single link (e.g.,
the member of the ``butterfly collection'' interest group who also belongs to
the ``origami'' interest group). Our definition of curvature implies
that such connecting
nodes tend to be hubs, and are characterized by a low curvature.

While the meta-in\-form\-ation of the graph
has been noted before \cite{12,10}, and used
for data mining in the influential works \cite{Kleinberg98,Brin98}, we
see now some differences with that work.
That approach views a link as conferring ``authority'' and searches the WWW for ``Authorities'' and ``Hubs'' that point at them. 
Based on these ideas the
successful search engine Google was designed; it also lead to the notion of
``Community'' \cite{Gibson98,NEC00}.
Similar investigations of the in\-form\-ation content of linkages are common in
other fields, such as studies of social \cite{FaustWasserman} or  paper
citation \cite{Small73} networks. However, in our case,
``authorities'' tend to have low curvature, while nodes recognized by
peers get high curvature. (Belonging to two groups lowers the
curvature by about a factor 2.)

\begin{table*}
\begin{center}
\begin{tabular}
{|l||r|r|r|r|r|}
\hline\hline
Anchor & URL encountered & URL read in & Clusters & Total links &
 Co-Links \\\hline\hline
Shakespeare & 277,114 &  69,982 & 1,560 &  3,730
& 321 \\\hline
 Needlework
&  341,398 &  102,895 & 1,498 &
4,440 & 727 \\\hline Revisionist & 66,771 & 22,933 &
 \ 947 & 1,307 & 67\\\hline Piazzolla & 47,978 &
 20,404 & 2,125 &
2,577 &70 \\\hline Mol. Biol. & 318,705 &  110,286 & 1,518
&
6,351 &868\\
\hline\hline
\end{tabular}
\caption{Data collected in 5
experimental Web crawls. Our protocol ensures that nodes within distance $r=1$
of the anchor are fully explored to a given depth
$d=3$.  The crawls started at sites of literary studies of
Shakespeare,
instructions for needlework, a notorious Swiss revisionist site,
tango (specifically the music of Piazzolla), and a site for molecular
biology. The link counts are for the clustered graphs.}
\end{center}
\end{table*}
\startfigure
\centerline{\psfig{figure=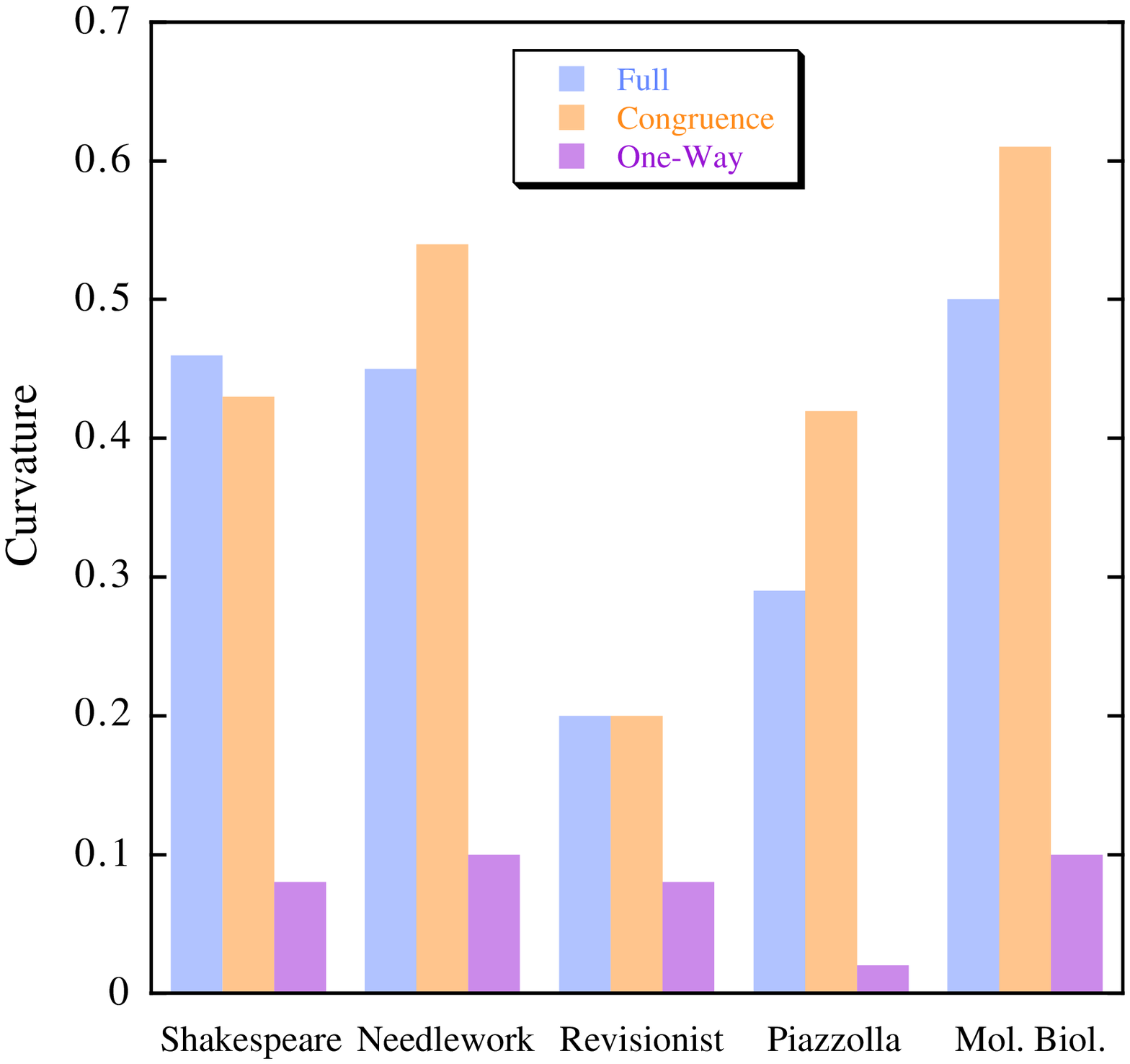,width=7cm}}
\caption{Curvature for the data of Table 1. The quantities are the
mean of the curvature of the anchor and its nearest neighbors. Full:
All links, irrespective of direction. Congruence: Only triangles with
3 co-links. One-way: Only triangles {\em without} co-links. Note that
the curvature is carried basically only by triangles with at least one
co-link.} 
\end{figure}
We tested our general ideas in several Web crawls, encompassing over one million URL's of which over three hundred thousand  were actually accessed. The database is stored for offline analysis, while intermediate results are used to direct the further action of the robot. Large numbers of links in a page (indicating the page is a hub) tend to stall the crawl, and we followed at most 35 links per page, neither were {\tt .cgi, .jpg} explored. As an aside, we note that {\tt .com}~sites seem to make little contribution to ``cooperative content'' and connect badly to others.

As the crawl proceeds, clustering according to Fig.~1 is implemented. To that end, links are determined to be {\em local} or {\em remote}. A local link is basically one inside a given site and any other link is remote. This decision is usually made by looking at the URL and using the known conventions for naming of pages inside a home directory (details can be obtained from the authors at {\tt http://mpej.unige.ch/\char`~eckmann}). The algorithm explores a page it encounters by following the local links down to depth $d$ (and perhaps $u$ upwards, links like ``back to home page'').

Crawls are initialized by choosing a first site (the {\em anchor}) and fetching it.  The links in a fetched page are identified and entered into the database. Our robot then fetches all the pages that these links point to.
After exhausting the search down and up the database is clustered and the robot is sent out along the first remote link. It explores that node in a similar way, going down and up the newly found local links as before. Once that is done the robot returns to the first node and follows the next remote link, until all remote links at distance $r=1$ from the anchor have been followed. The robot can then be directed to explore $r=2$ or sequentially larger distances, if needed.

The breadth of crawls is shown in Table 1, while results from specific
crawls are shown in Figure 3. The curvature was calculated for all
nodes within a distance $r=1$ of the anchor, and for three types of
graphs. The `Full' graph includes all links that were found in the
crawl, the `Congruence' graph is composed solely of co-links (the red
link of Fig.~1), and the `One-way' subgraph is obtained from the full
graph by removing all co-links (leaving only the black links). 

Our main result, shown in Fig.~4, is that the average curvatures of the
Congruence and Full graphs are very similar, while that of the One-way graph is
lower by a factor ranging between $2.5$ and $21$. This means that removal of
the co-links results in a dramatic decrease of the curvature. {\bf The
curvature of the full graph is carried predominantly by triangles containing at
least one reciprocal side}. This curvature is not produced by triangles with
three co-links alone, because of their small relative number, but rather from
triangles that have a combination of co- and One-way links. To put the results
of Fig.~3 in perspective, we note that the density of triangles (and
average curvature as well) of a random graph tends to zero with the size of the
graph \cite{Bollobas}.

For all experiments, the distribution of curvatures $c$ among the nodes of the
Full graphs obeys a power law, $c= (2.0 \pm 0.2) v^{-1}$, where $v$ is the
valence of the node. While the scaling is striking, we are not certain of its
origin. It does cause highly connected nodes with high $v$ such as hubs and
authorities to have, on average, a low curvature. This agrees with our
intuition that such nodes contribute perhaps to global connectivity, but not to
the local interest groups that curvature identifies.

While we consistently avoided any reference to content of sites in the crawl
itself, to check our method a separate, objective criterion was needed.  We
therefore checked manually whether the co-links reveal contextual linkages, by
reading in 784 nodes from our crawls (in most cases checking the name of the
node sufficed). Indeed, we found that no less than $75\%$ to $100\%$ of
nodes congruent to the anchor indeed relate to the same topic.

Going beyond the WWW, we examined the efficiency of curvature as a measure of
thematic cohesion by studying three other networks. These are the
neuronal network of the nematode {\it C. elegans}, the protein-protein
interaction network of the yeast {\it S. cerevisiae} and the citation network
of the mathematical-physics archive {\tt http://www.ma.utexas.edu/mp\char`_arc/}. In
each case we used an existing database to access and reconstruct the network,
and as a control each had a separate, objective criterion by which related
nodes were grouped into ``themes''.

Our analysis of the neural network of the worm relies on classic
work\cite{White} that used about 8,000 electron microscopy sections to trace
directed neuronal connections\cite{neurodata}. The control classification is
the currently accepted association of neurons with organs found by both
morphological and connective data \cite{White,Wood}, where the existence of
triangles was indeed noted, though only incidentally.

%\startfigure
\begin{figure*}
%\centerline{\psfig{figure=worm.ps,width=\columnwidth,angle=90}}
\centerline{\psfig{figure=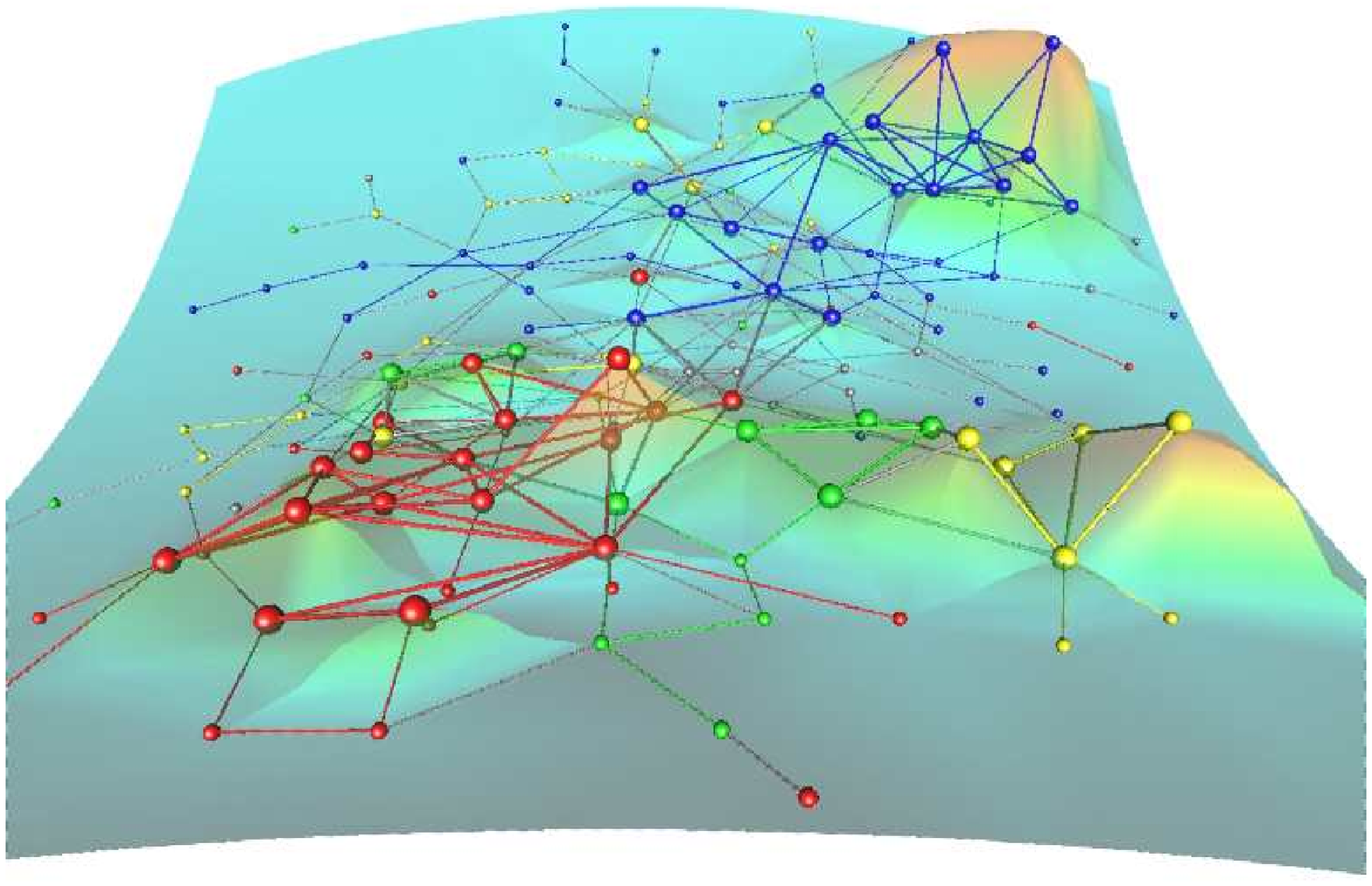,width=\textwidth,angle=0}}
%\centerline{\psfig{figure=ww3.ps,width=\columnwidth,angle=0}}
\caption{The worm brain. The height is proportional to curvature. The
red nodes are amphid cells, the yellow are other sensory neurons of the head, and blue are motor neurons of the nerve
ring. Only co-links are shown, and triangles are enhanced.} 
\end{figure*}

In Fig.~5 we present the network of the brain of the nematode. We
immediately note three connected components of high curvature, color
coded to display the association with the major neural circuits, or
organs. The elevated groups are the amphids, the motor neurons of the
nerve ring and ``other'' sensory neurons of the head. The other
circuits are either low-curvature (egg-laying and motor neurons of the
ventral cord) or absent from the compilation \cite{neurodata} (tail
ganglia). The amphids and nerve ring are particularly well separated
into connected components of the congruence graph. Furthermore, we
find that interneurons (CPU-like neurons) carry more curvature than
sensory (input) or motor (output) neurons, which agrees with our
intuition of how connections should distribute within the brain. Most
co-links ($61$\%) connect neurons that are within the same functional
circuit ($72$\% for co-links within triangles) while most of the
One-way links ($58$\%) connect between
two different organs. 

For the protein-protein interactions of the yeast we took the
protein-protein interaction database {\tt
http://dip.doe-mbi.ucla.edu\discretionary{}{}{}/dip\discretionary{}{}{}/Down\discretionary{}{}{}lo\discretionary{}{}{}a\discretionary{}{}{}d\discretionary{}{}{}.\discretionary{}{}{}cgi}, 
treating any interaction as a co-link (i.e., there are no One-way
links). The control was taken from the functional classification given
in the Yeast Protein Databank (YPD)  at {\tt
http://www.proteome.com\discretionary{}{}{}/databases\discretionary{}{}{}/YPD\discretionary{}{}{}/YPDcategories\discretionary{}{}{}/Functional\char`_\discretionary{}{}{}Cate\discretionary{}{}{}gorie\discretionary{}{}{}s\discretionary{}{}{}.html}. We found that proteins within the same connected groups of high
curvature almost always ($77$\% of the cases) belonged to the same
functional grouping of the YPD. At $c_n\ge 0.15$ there are $34$ such
groups, that include between $3$ to $18$ members, with an average of 5
proteins per group, (data not shown). 

In the case of the Mathematical Physics archive, we used the archive {\tt
http://www.ma.utexas.edu\discretionary{}{}{}/mp\char`_arc} to define a
network of citations between  authors in the same field.  A co-link
exists if author A cites author B (in any paper within the archive)
and author B cites A in any paper of his, while co-authors are
automatically co-linked. Note that our definition of co-links is
fundamentally different from the co-citations defined in
\cite{Small73}, since there no reciprocal recognition exists. The
control we used here was more problematic, relying on scanning
manually all conference pages listed at the International Mathematical
Physics Association homepage, and including an author in a group
(e.g., ``Quantum mechanics'', ``Field theory'' or ``Statistical
mechanics'') if she spoke under that topic in any of the
conferences. However, we were able to classify only about $400$ of the
$1700$ authors (since not every author is talking at these
conferences). As before, high curvature selected out the groups we
found manually, with $5$ of the $20$ or so topic groups we defined
very strongly represented and clearly identified (data not
shown). Again, ``authorities'', which in this case are very polyvalent
scientists, have lower curvature than specialists in a small field,
and serve rather as hubs. 

Our discovery opens a number of new possibilities for studying highly assembled
complex structures in general, perhaps even the brain. First, the introduction
of geometry frees us from the subjectivity of contextual concepts, such as
meaning, content, and the like. The locality develops a different direction
than global notions like scaling in the WWW and the ``small-world'' effect.
Second, the concept of congruence shows the usefulness of combining geometry
with selective action in probing sophisticated structures, including {\it C.
elegans}. In particular, we have seen how a very simple dynamical rule -- the
search for congruence -- radically changes our global view of the Web,
replacing statistical aspects with unsuspected relations which can be found
only by looking at combined, independently built, pathways in the Web. Our
approach can certainly be refined and extended in the Web, but should prove
useful in many other contexts.

Our study further shows that robots perceive the Web differently than the
people
who actually write and use the pages. Our robots go back and forth between
various sites, gaining a more coherent view of their relations. We have shown
that the geometric properties of the space in which they roam and the landscape
that they reconstruct reveal new connective meta-in\-form\-ation, hidden from
the
common user.

What can one expect in the future? Other forms of meta-in\-form\-ation, not
necessarily arising from connectivity, will certainly be found in the WWW and
other complex networks. The dynamics of developing interest groups is also an
important issue \cite{2,6}~ that may involve rapid fusion
processes \cite{Derrida}. {}From a mathematical perspective, we have
demonstrated
the need for concepts of local curvature in graphs. Our definition of curvature
seems a useful beginning to elicit Web properties, and is easily generalized to
balls of radius $r=2$ (next-nearest neighbors) or more, as well as to simplices
like tetrahedra. But more advanced geometric concepts, measuring
non-commutativity in the order of visiting sites, might be needed in the
future.

\normalsize

\thanks{\small We are grateful for numerous discussions with our
colleagues in
Geneva and Rehovot,
and for special help by A. Haefliger and D. Ruelle. We also thank
P. Blekken for help with Coin. This work was partially
supported by
the Fonds
National Suisse and the Minerva Foundation, Munich.}


\begin{thebibliography}{99}
\bibitem{Cluster}Note this is a different kind of cluster than that described
by the
``clustering
coefficient'' of
Watts and Strogatz.
\end{thebibliography}
\end{document}